\documentclass[11pt]{article}

\usepackage{amssymb}

\usepackage{amstext}
\usepackage{amsfonts}
\usepackage{amsmath}
\usepackage{graphicx}
\usepackage[letterpaper]{geometry}

\newtheorem{theorem}{Theorem}[section]    
\newtheorem{fact}[theorem]{Fact}    
\newtheorem{result}[theorem]{Result}    
\newtheorem{definition}{Definition}[section] 
\newtheorem{lemma}[theorem]{Lemma}    
\newcommand{\qed}{\hfill{$\rule{6pt}{6pt}$}} 
\newenvironment{proof}{\noindent{\bf Proof}:}{\qed}
\newenvironment{proofof}[1]{\noindent{\bf Proof of #1:}}{\qed}

\newcommand{\reals}{{\mathbb R}}
\newcommand{\size}[1]{\left| #1 \right|}
\newcommand{\set}[1]{\left\{ #1 \right\}}
\newcommand{\floor}[1]{\left\lfloor #1 \right\rfloor}
\newcommand{\re}{{\mathrm{e}}}
\newcommand{\eqdef}{\stackrel{\mathrm{def}}{=}}
\newcommand{\Tr}{{\mathrm{Tr}}}

\newcommand{\suppress}[1]{}

\newcommand{\E}{{\mathcal{E}}}

\newcommand{\cX}{{\mathcal{X}}}
\newcommand{\cY}{{\mathcal{Y}}}
\newcommand{\cZ}{{\mathcal{Z}}}
\newcommand{\rS}{{\mathrm{S}}}
\newcommand{\rD}{{\mathrm{D}}}
\newcommand{\rH}{{\mathrm{H}}}

\newcommand{\rU}{{\mathrm{U}}}
\newcommand{\rL}{{\mathrm{L}}}
\newcommand{\rI}{{\mathrm{I}}}
\newcommand{\QSC}{{\mathbf{QSC}}}
\newcommand{\bE}{{\mathbb{E}}}

\title{ {\bf A Separation between Divergence and Holevo Information
    for Ensembles}
}

\author{
Rahul Jain
\thanks{School of Computer Science, and
Institute for Quantum Computing,
University of Waterloo, 200 University Ave.\ W., Waterloo, ON N2L 3G1,
Canada. Email: {\tt rjain@cs.uwaterloo.ca}.
Research supported in part by ARO/NSA USA.
}\\
U.\ Waterloo
\and
Ashwin Nayak
\thanks{
Department of Combinatorics and Optimization, and Institute for
Quantum Computing, University of Waterloo, 200 University Ave.\ W.,
Waterloo, ON N2L 3G1, Canada.
E-mail: {\tt anayak@math.uwaterloo.ca}.
Research supported in part by NSERC Canada, CIFAR, MITACS, QuantumWorks,
and an ERA from the Province of Ontario.
A.N.\ is also Associate Member, Perimeter Institute for
Theoretical Physics, Waterloo, Canada.
Research at Perimeter Institute for Theoretical Physics is supported
in part by the Government of Canada through NSERC and by the Province
of Ontario through MRI.
}\\
U.\ Waterloo \& Perimeter
\and
Yi Su 
\thanks{
Department of Pure Mathematics,
University of Waterloo, 200 University Ave.\ W.,
Waterloo, ON N2L 3G1, Canada.
E-mail: {\tt y6su@student.math.uwaterloo.ca}.
Research supported in part by an NSERC Canada Undergraduate Research Award.
} \\
U.\ Waterloo
}

\date{December 5, 2007}

\begin{document}

\begin{titlepage}

\maketitle
\thispagestyle{empty}

\abstract{ The notion of {\em divergence information\/} of an ensemble
of probability distributions was introduced by Jain, Radhakrishnan,
and Sen~\cite{JainRS02,JainRS07} in the context of the ``substate
theorem''. Since then, divergence has been recognized as a more
natural measure of information in several situations in quantum and
classical communication. 

We construct ensembles of probability distributions for which
divergence information may be significantly smaller than the more
standard Holevo information. As a result, we establish that lower
bounds previously shown for Holevo information are weaker than similar
ones shown for divergence information.  }

\end{titlepage}

\section{Introduction}
\label{sec-intro}

In this article, we study the relationship between two different
measures of information contained in an ensemble of probability
distributions.  The first measure, {\em Holevo information\/}, is a
standard notion from information theory, and is equivalent to the
notion of {\em mutual information\/} between two random
variables. Consider jointly distributed random variables~$XY$,
with~$X$ taking values in a sample space~$\cX$. Consider the ensemble
of distributions~$\E = \set{ (\lambda_i,Y_i) \;:\; i \in \cX}$,
where $\lambda_i = \Pr(X = i)$, and~$Y_i = Y|(X = i)$, obtained by
conditioning on values assumed by~$X$. The Holevo information of the
ensemble is given by~$\chi(\E) = \rI(X:Y) = \bE_{i \sim X}
\rS(Y_i\|Y)$, where~$\rS(\cdot\|\cdot)$ measures the relative entropy
of a random variable (equivalently, distribution) with respect to
another.
This notion may be extended to ensembles of quantum states (see, e.g.,
the text~\cite{NielsenC00}), and the term `Holevo information' is
derived from the literature in quantum information theory.

The second measure, {\em divergence information\/}, was introduced by
Jain, Radhakrishnan, and Sen~\cite{JainRS02,JainRS07}. It arises in
the study of relative entropy, and its connection with a ``substate
property''. The {\em observational divergence\/} of two classical
distributions~$P,Q$ on the same finite sample space is~$\max_E P(E)
\log_2 (P(E)/Q(E))$, where~$E$ ranges over all events.  We may view
this as a (scaled) measure of the factor by which~$P$ may exceed~$Q$
for an event of interest.  The notion of {\em divergence
information\/} is derived from this as~$\rD(\E) = \bE_{i \sim X}
\rD(Y_i\|Y)$, in analogy with Holevo information. A quantum generalisation
of this measure may also be defined~\cite{JainRS07}.

Relative entropy and Holevo (or mutual) information have been studied
extensively in communication theory and beyond (see, e.g,
\cite{CoverT91}) as they arise in a variety of applications. Since the
discovery of the substate theorem~\cite{JainRS02}, divergence is being
recognized as a more natural measure of information in a growing
number of applications~\cite[Section~1]{JainRS07}. The applications
include privacy trade-offs in communicatioin protocols for computing
relations~\cite{JainRS05} and {\em bit-string
commitment\/}~\cite{Jain05}, and the communication complexity of {\em
remote state preparation\/}~\cite{Jain06}. In particular, divergence captures, up to
a constant factor, the substate property for probability
distributions. It thus becomes relevant in every application where the
substate theorem is used.

We construct ensembles of probability distributions (equivalently,
jointly distributed random variables) for which the Holevo and
divergence information are quantitatively different.
\begin{theorem}
\label{thm-intro}
For every positive integer~$N$, and real number~$k$ such that~$N >
2^{36k^2}$, there is an ensemble~$\E$ of distributions over a sample
space of size~$N$ such that~$\rD(\E) = k$ and~$\chi(\E) =
\Theta(k\log\log N)$.
\end{theorem}
A more precise statement of this theorem (Theorem~\ref{thm-ensemble}) 
and related results may be found in Section~\ref{sec-ensemble}.

The ensembles we construct satisfy the property that the ensemble
average (i.e., the distribution of the random variable~$Y$ in the
description above) is uniform. We show that the above separation is
essentially the best possible whenever the ensemble
average is uniform (Theorem~\ref{thm-ub}).
The result also applies to ensembles of quantum
states, where the ensemble average is the completely mixed state
(Theorem~\ref{thm-quantum-ub}). We
leave open the possibility of larger separations for classical or quantum
ensembles with non-uniform averages.

The difference between the two measures demonstrated by
Theorem~\ref{thm-intro} shows that in certain applications, divergence
is quantitatively a more relevant measure of information. In
Section~\ref{sec-impl}, we describe two applications where
functionally similar lower bounds have been established in terms of
both measures. This article shows that the lower bounds in terms
of divergence information are, in fact, stronger.

In prior work on the subject, Jain {\it et
al.\/}~\cite[Appendix~A]{JainRS07} compare relative entropy and
divergence for classical as well as quantum states.
For pairs of distributions~$P,Q$ over a sample space of
size~$N$, they show that~$\rD(P\|Q) \le \rS(P\|Q) + 1$, and~$\rS(P\|Q)
\le \rD(P\|Q) \cdot (N-1)$. This extends to the corresponding measures
of information in an ensemble: $\rD(\E) \le \chi(\E) + 1$ and~$\chi(\E)
\le \rD(\E) \cdot (N-1)$. They show qualitatively similar relations for
ensembles of
quantum states.  In addition, they construct a pair of
distributions~$P,Q$ such that~$\rS(P\|Q) = \Theta(\rD(P\|Q) \cdot
N)$. However, their construction does not appear to translate to a
similar separation for {\em ensembles\/} of probability
distributions. Our work fills this gap for ensembles (of classical or
quantum states) with a uniform average.

\section{Preliminaries}         
\label{sec-prelims}

Here, we summarise our notation and the information-theoretic concepts
we encounter in this work. We refer the reader to the text by Cover and
Thomas~\cite{CoverT91} for a deeper treatment of (classical) information
theory. While the bulk of this article pertains to classical information
theory, as mentioned in Section~\ref{sec-intro}, it is motivated by studies
in (and has implications for) quantum information. We refer the
reader to the text~\cite{NielsenC00} for an introduction to quantum
information.

For a positive integer $N$, let $[N]$ represent the set $\{1, \ldots,
N \}$. We view probability distributions over~$[N]$ as vectors
in~$\reals^N$. The probability assigned by distribution~$P$ to a
sample point~$i \in [N]$ is denoted by~$p_{i}$ (i.e., with the same
letter in small case). We denote by~$P^\downarrow$ the distribution
obtained from~$P$ by composing it with a permutation~$\pi$ on~$[N]$ so
that~$p^\downarrow_i = p_{\pi(i)}$ and~$p^\downarrow_1 \ge
p^\downarrow_2 \ge \cdots \ge p^\downarrow_N$.  For an event~$E
\subseteq [N]$, let~$P(E) = \sum_{i \in E} p_i$ denote the probability
of that event. We denote the uniform distribution over~$[N]$
by~$\rU_N$.

We appeal to the {\em majorisation\/} relation for some of our
arguments. The relation tells us which of two given distributions is
``more random''.
\begin{definition}[Majorisation]
Let~$P,Q$ be distributions over~$[N]$. We say that~$P$ {\em
majorises\/}~$Q$, denoted as $P\succeq Q$, if
\begin{eqnarray*}
\sum_{j=1}^{i} p^\downarrow_{j} & \ge &  \sum_{j=1}^{i}
q^\downarrow_{j},
\end{eqnarray*}
for all~$i \in [N]$.
\end{definition}
The following is straightforward.
\begin{fact}
\label{fact-maj-uniform}
Any probability distribution~$P$ on~$[N]$ majorises~$\rU_N$, the
uniform distribution over~$[N]$.
\end{fact}

Throughout this article, we use `$\log$' to denote the logarithm with
base~$2$, and~`$\ln$' to denote the logarithm with base~$\re$.
\begin{definition}[Entropy, relative entropy] 
\label{def-relative-entropy}
Let $P,Q$ be probability distributions on $[N]$. The {\em entropy\/}
of~$P$ is defined as~$\rH(P) \eqdef - \sum_{i = 1}^N p_i \log p_i$.
The {\em relative entropy\/} between~$P,Q$, denoted $\rS(P \| Q)$, is
defined as
\[ 
\rS(P\|Q) \quad \eqdef \quad \sum_{i=1}^{N}p_{i}\log \frac{p_{i}}{q_{i}}
~.
\]
\end{definition}
Note that the relative entropy with respect to the uniform
distribution is connected to entropy as $\rS(P\|\rU_N) = \log N -
\rH(P)$.

We can formalise the connection between majorisation and randomness
through the following fact.
\begin{fact}
\label{fact-maj-entropy}
If~$P,Q$ are distributions over~$[N]$ such that~$P$ majorises~$Q$,
i.e.~$P \succeq Q$, then $\rH(P) \leq \rH(Q)$.
\end{fact}

The notion of {\em observational divergence\/} was defined by Jain,
Radhakrishnan, and Sen~\cite{JainRS02} in the context of the
``substate theorem''.
\begin{definition}[Observational divergence]
\label{def-div}
Let $P,Q$ be probability distributions on $[N]$. Then the {\em
observational divergence\/} between them, denoted $\rD(P \| Q)$, is
defined as
\[
\rD(P\|Q)\quad \eqdef \quad \max_{E \subseteq [N]} \; P(E)\log
    \frac{P(E)}{Q(E)} ~.
\]
\end{definition}
Throughout the paper we refer to `observational divergence' as simply
`divergence'.

Divergence is always non-negative, and the divergence of any
distribution with respect to the uniform distribution is bounded.
\begin{lemma}
\label{lem-div-uniform}
For any probability distribution~$P$ on~$[N]$, we have~$0 \le
\rD(P\|\rU_N) \le \log N$.
\end{lemma}
\begin{proof}
  Consider the event~$E$ which achieves the divergence between~$P$
  and~$\rU_N$.  W.l.o.g., the event~$E$ is non-empty. Therefore~$P(E)
  \ge \rU_N(E) \ge 1/N$, and~$0 \le \rD(P\|U_N) \le P(E) \log P(E)N \le \log N$.
\end{proof}

We observe that we need only maximise over~$N$ events to calculate
divergence with respect to the uniform distribution.
\begin{lemma}
\label{lem-div-def}
For any probability distribution~$P$ on~$[N]$ such that~$P^\downarrow
= P$, i.e., $p_1 \ge p_2 \ge \cdots \ge p_N$, we have
\[
\rD(P\| \rU_N) \quad = \quad \max_{i \in [N]} \; P([i])\log \frac{N \cdot
  P([i])}{i} ~.
\]
\end{lemma}
\begin{proof}
  By definition of observational divergence, the RHS above is bounded
  by~$\rD(P\| \rU_N)$. For the inequality in the other direction, we
  note that the probability~$P(E)$ of any event~$E$ with size~$n_E =
  \size{E}$ is bounded by~$P([n_{E}])$, the probability of the
  first~$n_{E}$ elements in~$[N]$.  We thus have
\begin{eqnarray*}
\rD(P\|Q) 
    & = & \max_{E \subseteq [N]} P(E) \log \frac{N \cdot P(E)}{n_E} \\
    & \le & \max_{E \subseteq [N]} P(E) 
          \log \frac{N \cdot P([n_E])}{n_E} \\
    & \le & \max_{E \subseteq [N]}  P([n_E]) 
          \log \frac{N \cdot P([n_E])}{n_E},
\end{eqnarray*}
since~$P$ majorises $\rU_N$ (Fact~\ref{fact-maj-uniform})
and~$P([n_E]) \ge \tfrac{n_E}{N}$.  This is equivalent to the RHS in
the statement of the lemma.
\end{proof}

\begin{definition}[Ensemble]
  \label{def-ens}
  An {\em ensemble\/} is a sequence of pairs~$\set{
    (\lambda_{j},Q_{j}) \;:\; j \in [M]}$, for some integer~$M$, where
  Let $\Lambda = (\lambda_j) \in \reals^M$ is a probability
  distribution on~$[M]$ and~$Q_{j}$ are probability distributions over
  the same sample space.
\end{definition}

\begin{definition}[Holevo information] 
\label{def-chi}
The {\em Holevo information\/} of an ensemble~$\E = \{(p_{j}, Q_{j})
\;:\; j\in[M]\}$, denoted as $\chi(\E)$,  is defined as
\[ 
\chi(\E) \quad \eqdef \quad \sum_{j=1}^{M}
\lambda_{j} \,\rS(Q_{j}\|Q), 
\] 
where $Q= \sum_{j=1}^{M} \lambda_{j} Q_{j}$ is the {\em ensemble
  average\/}.
\end{definition}

\begin{definition}[Divergence information]
\label{def-divinfo}
The {\em divergence information\/} of an ensemble~$\E = \{(p_{j},
Q_{j}) \;:\; j\in[M]\}$, denoted as $\rD(\E)$ is defined as
\[  
\rD(\E) \quad \eqdef \quad \sum_{j=1}^{M} \lambda_{j} \,\rD(Q_{j}\|Q),
\] 
where $Q= \sum_{j=1}^{M} \lambda_{j} Q_{j}$ is the {\em ensemble
  average\/}.
\end{definition}

\section{Divergence versus relative entropy}
\label{sec-ensemble}

In this section, we describe the construction of an ensemble for which
there is a large separation between divergence and Holevo
information. The ensemble has the property that the ensemble average
is uniform. As a by-product of our construction, we also obtain a
bound on the maximum possible separation for ensembles with a uniform
average.

We begin with the construction of the ensemble.  Let~$f_\rL(k,N) =
k(\ln \log (kN) - \ln (6k) +1 ) - \log(1 + k\ln 2) - 1 - \frac{1}{\ln 2}$
on point in the positive orthant in~$\reals^2$ with~$Nk > 1$.
\begin{theorem}
\label{thm-ensemble}
For every integer~$N > 1$, and every positive real
number~$\tfrac{16}{N} \le k < \log N$, there is an ensemble~$\E =
\set{ (\tfrac{1}{N}, Q_i) \;:\; i \in [N] }$ with~$\tfrac{1}{N} \sum_i
Q_i = \rU_N$, the uniform distribution over~$[N]$, with~$\rD(\E) \le
k$, and
\[
\chi(\E) \quad \geq \quad f_\rL(k,N).
\]
\end{theorem}

To construct the ensemble described in the theorem above, we first
construct a probability distribution~$P$ on~$[N]$ with observational
divergence~$\rD(P\| \rU_N) \le k$ such that its relative entropy~$\rS(P\|
\rU_N)$ is large as compared with~$k$. Let~$f_\rU = k(\ln \log(Nk) -
\ln k + 1)$ be defined on points in the positive orthant of~$\reals^2$
with~$kN > 1$.
\begin{theorem}
\label{thm-distribution}
For every integer~$N > 1$, and every positive real
number~$\tfrac{16}{N} \le k < \log N$, there is a probability
distribution~$P$ with~$\rD(P\| \rU_N) = k$, and
\[
f_\rL(k,N) \quad \le \quad \rS(P\| \rU_N) \quad \le \quad f_\rU(k,N).
\]
\end{theorem}
The construction of the ensemble is now immediate.

\begin{proofof}{Theorem~\ref{thm-ensemble}}
  Let~$Q_j = P \circ \pi_j$, where~$\pi_j$ is the cyclic permutation
  of~$[N]$ by~$j-1$ places.  We endow the set of the~$N$ cyclic
  permutations~$\set{Q_j \;:\; j \in [N]}$ of~$P$ with the uniform
  distribution. By construction, the ensemble average
  is~$\rU_N$. Since both observational divergence and relative entropy
  with respect to the uniform distribution are invariant under
  permutations of the sample space, $\rD(\E) = \rD(P\|\rU_N) \le k$,
  and~$\chi(\E) = \rS(P\| \rU_N) \geq f_\rL(k,N)$.
\end{proofof}

We turn to the construction of the distribution~$P$. Our construction
is such that~$P^\downarrow = P$, i.e., $p_1 \ge p_2 \ge \cdots \ge
p_N$. Lemma~\ref{lem-div-def} tells us that we need only ensure that
\begin{eqnarray}
\label{eqn-div-constraint}
P([i]) \log \frac{N \cdot P([i])}{i} 
    & \le & k, \quad \forall~i \in [N],
\end{eqnarray}
to ensure~$\rD(P\| Q) \le k$.  Since~$\rS(P\|\rU_N) = \log N -
\rH(P)$, we wish to minimise the entropy of~$P$ subject to the
constraints in Eq.~(\ref{eqn-div-constraint}). This is equivalent to
successively maximising~$p_1, p_2, \ldots$, and motivates the
following definitions.

Define the function~$g(y,x) = y \log (Ny/x) - k$ on the positive
orthant of~$\reals^2$. Consider the function~$h : \reals^+ \rightarrow
\reals^+$ implicitly defined by the equation~$g(h(x),x) = 0$. 
\begin{lemma}
\label{lem-well-def}
The function~$h \;:\; \reals^+ \rightarrow \reals^+$ is well-defined,
strictly increasing, and concave.
\end{lemma}
\begin{proof}
  Fix an~$x \in \reals^+$, and consider the function~$g_x(y) =
  g(y,x)$.  This function is continuous on~$\reals^+$, tends to~$-k <
  0$ as~$y \rightarrow 0^+$, and tends to~$\infty$ as~$y \rightarrow
  \infty$. By Intermediate Value Theorem, for some~$y > 0$, we
  have~$g_x(y) = 0$.  Moreover, $g_x(y) < -k$ for~$0 < y \le x/N$, and is
  strictly increasing for~$y > x/N\re$ (its derivative is~$g'_x(y) =
  \log\tfrac{\re Ny}{x}$).  Therefore there is a unique~$y$ such
  that~$g_x(y) = 0$ and~$h(x)$ is well-defined.

The function~$h$ satisfies the equation~$h \log \tfrac{Nh}{x} = k$, and
therefore the identity
\[
x \quad = \quad Nh \exp\!\left( - \tfrac{k \ln 2}{h} \right).
\]
Differentiating with respect to~$h$, we see that
\begin{eqnarray*}
\frac{dx}{dh} & = & N \left( 1 + \frac{k \ln 2}{h} \right)
                    \exp\!\left( - \tfrac{k \ln 2}{h} \right), \text{ and}
                    \\
\frac{d^2 x}{dh^2} & = & \frac{N (k \ln 2)^2}{h^3} \;
                    \exp\!\left( - \tfrac{k \ln 2}{h} \right).
\end{eqnarray*}
So~$\tfrac{dh}{dx} > 0$ for all~$x > 0$, and~$h$ is a strictly increasing
function. Note also that~$\tfrac{d^2 x}{dh^2} > 0$ for all~$h > 0$,
so~$x$ is a convex function of~$h$. Since~$h$ is an increasing function,
convexity of~$x(h)$ implies concavity of~$h(x)$.
\end{proof}

Let~$v_0 = 0$.  For~$i \in [N]$, let~$v_i = h(i)$, i.e., $v_{i} \log
\tfrac{Nv_{i}}{i} = k$.  Let $s_{i} \eqdef \min\{1, v_{i}\}$, for~$i \in
[N]$. Let~$p_1 = s_1$, and~$p_{i} =s_{i}-s_{i-1}$ for all~$2 \le i \le
N$. Lemma~\ref{lem-well-def} guarantees that these numbers are
well-defined. We claim that
\begin{lemma}
\label{lem-prob-distr}
The vector~$P = (p_i) \in \reals^N$ defined above is a probability
distribution, and~$P^\downarrow = P$, i.e., $p_1 \ge p_2 \ge \cdots
\ge p_N$.
\end{lemma}
\begin{proof}
By definition, we have~$v_i > 0$ for all~$i \in [N]$. Therefore $s_1 =
\min \set{1, v_1} > 0$. Since~$h(x)$ is an increasing function in~$x$,
the sequence~$(v_i)$ is also increasing, so~$(s_i)$ is non-decreasing.
Therefore~$p_i = s_i - s_{i-1} \ge 0$ for~$i > 1$.

Now~$v_N \log v_N = k > 0$. Since~$x \log x \le 0$ for~$x \in (0,1)$,
we have~$v_N > 1$. So~$s_N = \min\set{1,v_N} = 1$.  Therefore~$\sum_{i
  = 1}^N p_i = s_N = 1$. So~$P$ is a probability distribution on~$[N]$.

Note that~$(v_2/2) \log( N v_2/2) = k/2 < k$, so~$v_1 >
v_2/2$. So~$s_1 \ge s_2/2 \Leftrightarrow p_1 \ge p_2$.  For~$i \ge
2$, we have~$p_i - p_{i+1} = (s_{i}-s_{i-1})-(s_{i+1}-s_{i})=2 s_{i} -
s_{i-1} - s_{i+1}$.  Since~$h(x)$ is concave, so is the
function~$\min\set{1,h(x)}$.  Therefore,~$s_{i} \geq
(s_{i-1}+s_{i+1})/2$, and the sequence~$(p_i)$ is non-decreasing.
\end{proof}

The vector~$S = (s_i) \in \reals^N$ thus represents the (cumulative)
distribution function corresponding to~$P$.

\begin{proofof}{Theorem~\ref{thm-distribution}}
    We claim that the probability distribution~$P$ constructed above
    satisfies the properties stated in the theorem.

    Since~$P^\downarrow = P$, by Lemma~\ref{lem-div-def}, we need only
    verify that~$s_i \log(N s_i/i) \le k$ for~$i \in [N]$. If~$s_i =
    v_i$, then the condition is satisfied with equality. (Note that
    since~$k < \log N$, we have~$s_1 = v_1 < 1$.) Else,~$s_i = 1 < v_i$,
    so~$s_i \log(N s_i/i) < v_i \log(N v_i/i) = k$.

    We now bound the relative entropy~$\rS(P\|\rU_N)$ from
    above. Let~$n$ be the smallest positive integer such that~$v_{n-1}
    \le 1$ and~$v_n > 1$. Note that~$n > 1$. We also have~$n \le N$,
    since~$v_N > 1$ (as $v_N \log v_N = k > 0$). Therefore, we
    have~$s_i = v_i$ (equivalently,~$N s_i =i2^{k/s_i}$) for~$i \in
    [n-1]$, and~$s_n = 1 < v_n$. Thus, for~$1 < i < n$,
\begin{eqnarray*}
  N p_{i}
  & = & i 2^{\frac{k}{s_{i}}} - (i-1)2^{\frac{k}{s_{i-1}}}  \\
  & = & 2^{\frac{k}{s_{i}}} + (i-1)(2^{\frac{k}{s_{i}}} 
        - 2^{\frac{k}{s_{i-1}}}) \\
  & = & 2^{\frac{k}{s_{i}}} + (i-1)2^{\frac{k}{s_{i-1}}}
        (2^{\frac{k}{s_{i}} - \frac{k}{s_{i-1}}}-1) \\
  & = & 2^{\frac{k}{s_{i}}} + N s_{i-1}
        (2^{\frac{k}{s_{i}} - \frac{k}{s_{i-1}}}-1) \\
  & \ge & 2^{\frac{k}{s_{i}}} + N s_{i-1} 
          \left( \frac{k}{s_{i}}-\frac{k}{s_{i-1}} \right) \ln2 \\
  & = & 2^{\frac{k}{s_{i}}} - \frac{Np_{i}k}{s_{i}}\ln2 .
\end{eqnarray*}
The penultimate line follows from the inequality~$2^x \ge 1 + x \ln 2$
for all~$x \in \reals$. Thus we have
\begin{eqnarray}
\label{eqn-lbp}
Np_{i} \quad \geq \quad \frac{2^{\frac{k}{s_{i}}}}{1+\frac{k}{s_{i}}\ln2}.
\end{eqnarray}
Since~$N p_1 = N s_1 = 2^{\frac{k}{s_1}}$, this also holds for~$i = 1$.

We bound the relative entropy using Eq.~(\ref{eqn-lbp}).
\begin{eqnarray}
\nonumber
\rS(P\|\rU_N) 
   & = & \sum_{i=1}^{N} p_{i} \log Np_{i} 
         \quad = \quad \sum_{i=1}^{n}p_{i}\log Np_{i} \\
\nonumber
   & \geq & \sum_{i=1}^{n-1} p_{i} 
            \log \frac{2^{\frac{k}{s_{i}}}}{1+\frac{k}{s_{i}}\ln2}
            + p_{n} \log N p_{n} \\
\label{eqn-rel-terms}
   & \geq &  \sum_{i=1}^{n-1} \frac{p_{i}k}{s_{i}} 
             - \sum_{i=1}^{n-1} p_{i} 
             \log\!\left(1 + \frac{k\ln2}{s_{i}} \right) 
             + p_{n} \log Np_{n}.
\end{eqnarray}
We bound each of the three terms in the RHS of
Eq.~(\ref{eqn-rel-terms}) separately.

We start with~$\sum_{i=1}^{n-1} \frac{p_{i}k}{s_{i}}$.  Let $p = p_1$,
and let $m = \floor{\tfrac{1}{p}}$. For every~$j \in [m]$, there is
an~$i \in [n]$, say~$i = i_j$, such that~$jp \le s_{i_j} \le
(j+1)p$. (Otherwise, for some~$i > 1$, the probability~$p_i = s_i -
s_{i-1}$ is strictly larger than~$p$, an impossibility.)

\begin{figure}[h]
\begin{center}
\includegraphics[scale=0.5]{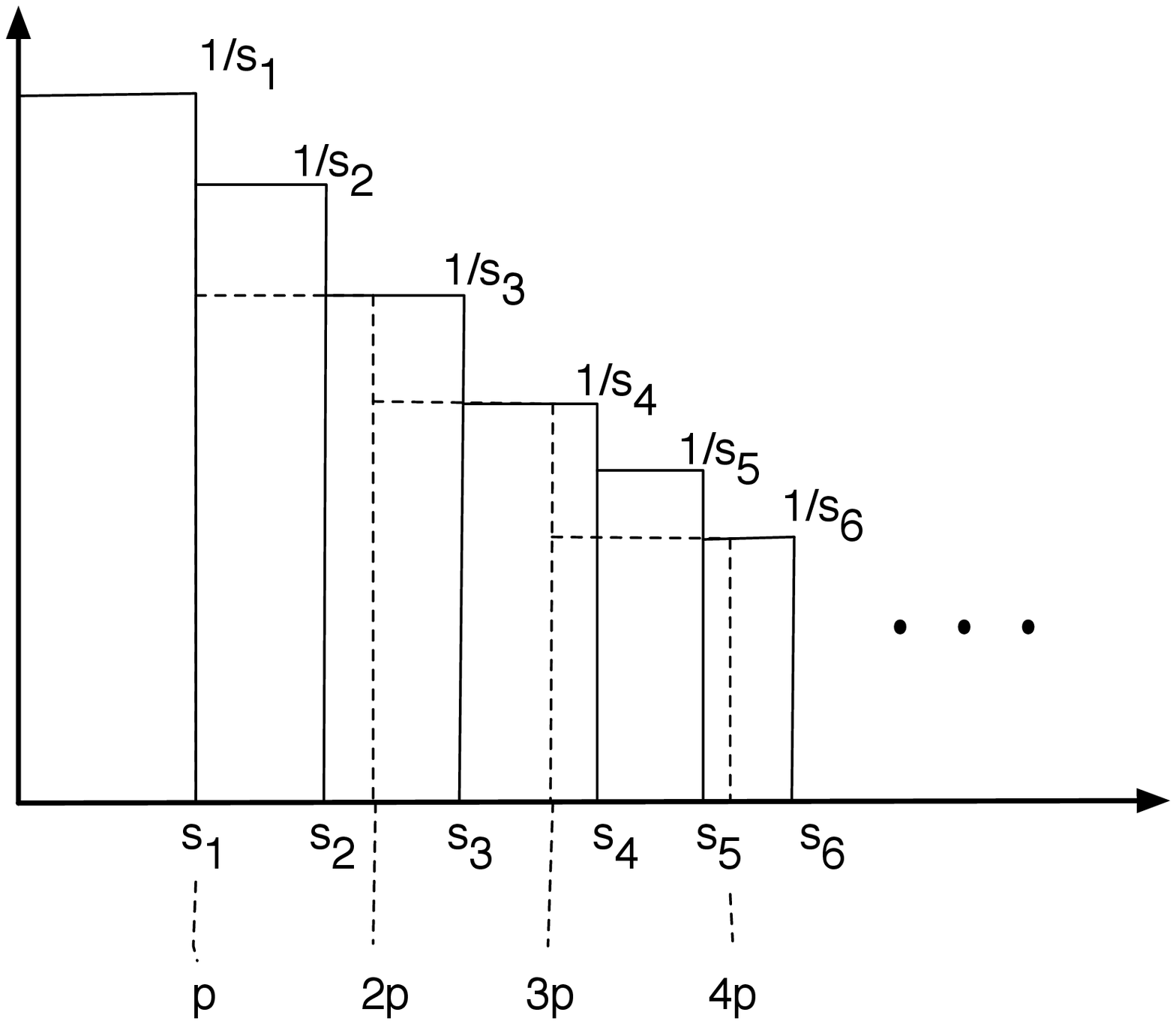}
\label{fig1}
\end{center}
\end{figure}

We interpret the sum~$ \sum_{i=2}^{n-1} \frac{p_{i}}{s_{i}} = \sum_{i
  = 2}^{n-1} \frac{s_i - s_{i-1}}{s_i}$ as a Riemann sum approximating
the area under the curve~$1/x$ between~$s_1$ and~$s_{n-1}$ with the
area under the solid lines in Figure~\ref{fig1}. This area is bounded
from below by the area under the dashed lines, which corresponds to
the area of rectangles of uniform width~$p$ and height~$1/s_{j+1}$ for
the~$j$th interval. Thus,
\begin{eqnarray}
\nonumber
\sum_{i=1}^{n-1} \frac{p_{i}k}{s_{i}}
    & \geq & k + k \sum_{j=1}^{m} p \cdot \frac{1}{s_{i_{j+1}}} \\
\nonumber
    & \geq & k + k \sum_{j=1}^{m} p \cdot \frac{1}{(j+2)p} \\
\nonumber
    & =    & k + k \sum_{j=1}^{m} \frac{1}{j+2} \\
\nonumber
    & \geq & k + k \int_3^{m+3} \frac{1}{x} dx \\
\label{eqn-term1-1}
    & = & k + k \ln \frac{m+3}{3}.
\end{eqnarray}

We lower bound $m = \floor{\tfrac{1}{p}}$ next. Recall that~$g_1(y) =
y \log(Ny)$ is an increasing function for~$y > \tfrac{1}{\re N}$,
and~$p = p_1 \ge 1/N$. Consider the value of~$g_1(y)$ at the point~$q
= \tfrac{2k}{\log kN}$:
\[
g_1(q) \quad = \quad \frac{2k}{\log kN}\log \frac{2Nk}{\log kN} 
\quad > \quad 2k \left(1 - \frac{\log \log kN}{\log kN}\right)
\quad \ge \quad k,
\]
since~$kN \ge 16$. As~$g_1(q) > g_1(p) > 0$, we have~$q >
p$. Therefore, $m \ge \tfrac{1}{p} - 1 \ge \tfrac{\log kN}{2k} -
1$. Together with Eq.~(\ref{eqn-term1-1}), we get
\begin{eqnarray}
\label{eqn-term1-lb}
\sum_{i=1}^{n-1} \frac{p_{i}k}{s_{i}} 
    & \ge & k (\ln \log kN - \ln 6k + 1).
\end{eqnarray}

Next, we derive a lower bound for the second term in
Eq.~(\ref{eqn-rel-terms}).
\begin{eqnarray}
\nonumber
- \sum_{i=1}^{n-1} p_{i} \log\!\left( 1+\frac{k\ln2}{s_{i}} \right)
    & = & - \sum_{i=1}^{n-1} p_{i} \log(s_{i}+k\ln2) 
          + \sum_{i=1}^{n-1}p_{i}\log s_{i} \\
\label{eqn-term2-1}
    & \geq & - \log (1+k\ln2) 
             + \sum_{i=1}^{n-1} p_{i} \log s_{i}.
\end{eqnarray}
Viewing the second term above as a Riemann sum, we get
\begin{eqnarray}
\nonumber
\sum_{i=1}^{n-1} p_{i}\log s_{i}
    & \geq & \int_{0}^{s_{n-1}}\log x \; dx \\
\nonumber
    & \geq & \int_{0}^{1} \log x \; dx \\
\label{eqn-term2-2}
    & = & -\frac{1}{\ln2}.
\end{eqnarray}
Combining Eq.~(\ref{eqn-term2-1}) and~(\ref{eqn-term2-2}), we get
\begin{eqnarray}
\label{eqn-term2-lb}
- \sum_{i=1}^{n-1} p_{i} \log\!\left(1+\frac{k\ln2}{s_{i}} \right)
    & \geq & -\log (1+k\ln2) -\frac{1}{\ln2}.
\end{eqnarray}

We bound the third term in Eq.~(\ref{eqn-rel-terms}) crudely as~$
p_{n}\log Np_{n} \geq -1$. Along with the bounds for the previous two terms,
Eq.~(\ref{eqn-term1-lb}), (\ref{eqn-term2-lb}),
this shows that
\begin{equation}
\rS(P\|\rU_N) \quad \ge \quad f_\rL(k,N)
\quad \eqdef \quad k(\ln\log kN-\ln6k +1)-\log (1+k\ln2) - 1 -\frac{1}{\ln2}.
\end{equation}
This proves the lower bound on the relative entropy. 

Moving to an upper bound, we have for~$i \ge 2$,
\begin{eqnarray*}
Np_{i} & = & i 2^{\frac{k}{s_{i}}}-(i-1)2^{\frac{k}{s_{i-1}}}  \\
       & = & 2^{\frac{k}{s_{i}}}+(i-1)(2^{\frac{k}{s_{i}}} 
             - 2^{\frac{k}{s_{i-1}}}) \\
       &\leq& 2^{\frac{k}{s_{i}}},
\end{eqnarray*}
since the second term is negative. This also holds for~$i = 1$,
since~$p_1 = s_1$ and~$s_1 \log Ns_1 = k$. Therefore,
\begin{eqnarray*}
\rS(P\| \rU_N) &=& \sum_{i=1}^{n}p_{i}\log Np_{i} \\
        &\leq& \sum_{i=1}^{n} \frac{kp_{i}}{s_{i}} \\
        &\leq& k + k \int_{s_{1}}^{1}\frac{1}{s}ds \\
        &=& k - k\ln s_{1} \\
        &\leq& k + k\ln\!\left(\frac{\log Nk}{k}\right) \\
        &=& k(1-\ln k+\ln(\log Nk)).
\end{eqnarray*}
In the last inequality, we used the lower bound~$s_1 \ge k/\log Nk$.
\end{proofof}

The upper and lower bounds on the relative entropy of~$P$ with respect
to the uniform distribution both behave as~$k\log\log Nk$ up to
constant factors.

\begin{proofof}{Theorem~\ref{thm-intro}}
The dominating term in both of lower bound and upper bound on the
relative entropy~$\rS(P\|\rU_N)$, is~$k \ln\log Nk$ when $N$ is large as
compared with~$k$. Specifically, when~$N > 2^{36k^2}$, we have
\[\frac{1}{2}k\log\log Nk \quad \leq \quad \rS(P\|\rU_N)
\quad \leq \quad 2 k\log\log Nk. \] Since~$k \le \log N$
(Lemma~\ref{lem-div-uniform}), $\rS(P\|\rU_N) = 
\Theta(\rD(P\|Q)\log\log N)$. The same holds for the ensembles
constructed in Theorem~\ref{thm-ensemble}.
\end{proofof}


The separation we demonstrated above is the best possible for
ensembles of distributions that have a uniform average distribution.
\begin{theorem}
\label{thm-ub}
For any positive integer~$N$, and any ensemble~$\E = \set{(\lambda_j,
  Q_j) \;:\; j \in [M]}$ of distributions over~$[N]$ such
that~$\sum_{j = 1}^M \lambda_j Q_j = \rU_N$, we have
\begin{eqnarray*}
  \chi(\E) & \le & K(2 \ln\log N - \ln K + 1) + 16,
\end{eqnarray*}
where~$K = \rD(\E)$.
\end{theorem}
\begin{proof}
  Let~$\rD(Q_j\|\rU_N) = k_j$. We show that~$\rS(Q_j\|\rU_N) \le k_j
  (2 \ln\log N - \ln k_j + 1)$ when~$k_j \ge \tfrac{16}{N}$.
  When~$k_j < \tfrac{16}{N}$, we have~$\rS(Q_j\|\rU_N) < 16$.
  Since~$k(2 \ln\log N - \ln k + 1)$ is a concave function in~$k$,
  averaging over~$j$ with respect to the distribution~$\Lambda =
  (\lambda_j)$ gives the claimed bound.

  Fix an~$j$ such that~$k_j > \tfrac{16}{N}$. Let~$R =
  Q_j^\downarrow$. Note that~$\rD(R\|\rU_N) = k_j$ and~$\rS(R\|\rU_N)
  = \rS(Q_j\|\rU_N)$. Consider the distribution~$P$ constructed as in
  Section~\ref{sec-ensemble} with~$k = k_j$. Using the notation of
  that section, we have~$s_i \log(N s_i/i) = k_j$ for all~$i < n$,
  and~$s_n = 1$. Let~$t_i = \sum_{l = 1}^i r_i$. By definition, we
  have~$t_i \log(N t_i/i) \le k_j = s_i \log(N s_i/i)$. Since the
  function~$g_i(y) = y \log(Ny/i)$ is strictly increasing for~$y \ge
  i/N\re$, and~$t_i \ge i/N$ (Fact~\ref{fact-maj-uniform}), we
  have~$t_i \leq s_i$ for~$i < n$. Since~$s_i = 1$ for~$i \geq n$, we
  have~$t_i \le s_i$ for these~$i$ as well. In other words, $P \succeq
  R$. By Fact~\ref{fact-maj-entropy}, $\rH(P) \leq \rH(R)
  \Leftrightarrow \rS(R\|\rU_N) \le \rS(P\|\rU_N)$. By
  Theorem~\ref{thm-distribution}, $\rS(P\|\rU_N) \leq k_j (\ln\log(N
  k_j) -\ln k_j + 1)$. Since~$k_j \le \log N$, this is at most~$k_j (2
  \ln\log N - \ln k_j +1)$.
\end{proof}

Finally, we observe that this is also the best separation possible for
an ensemble of quantum states with a completely mixed ensemble
average. 
\begin{theorem}
\label{thm-quantum-ub}
For any positive integer~$N$, and any ensemble~$\E = \set{(\lambda_j,
  \rho_j) \;:\; j \in [M]}$ of quantum states~$\rho_j$ over a Hilbert
space of dimension~$N$ such that~$\sum_{j = 1}^M \lambda_j \rho_j =
\tfrac{\rI}{N}$, the completely mixed state of dimension~$N$, we have
\begin{eqnarray*}
  \chi(\E) & \le & K(2 \ln\log N - \ln K + 1) + 16,
\end{eqnarray*}
where~$K = \rD(\E)$.
\end{theorem}
\begin{proof}
  Let~$Q_j$ be the probability distribution on~$[N]$ corresponding to
  the eigenvalues of~$\rho_j$. By definition of observational
  divergence for quantum states, $\rD(Q_j\|\rU_N) \le
  \rD(\rho_j\|\tfrac{\rI}{N})$. Further, we
  have~$\rS(\rho_j\|\tfrac{\rI}{N}) = \rS(Q_j\|\rU_N)$. We now apply
  the same reasoning as in the proof of Theorem~\ref{thm-ub}, note
  that the divergence of the ensemble~$\set{(\lambda_j,Q_j) \;:\; j
    \in [M]}$ is bounded by~$\rD(\E)$, and that the RHS in the
  statement is a non-decreasing function of~$K$. This gives us the
  stated bound. (Note that we do not need~$\sum_{j = 1}^M \lambda_j
  Q_j = \rU_N$ to use the reasoning in Theorem~\ref{thm-ub}.)
\end{proof}

\pagebreak

\bibliographystyle{plain}

\pagebreak

\appendix

\section{Implications for quantum protocols}
\label{sec-impl}

\subsection{Quantum string commitment}
\label{sec-qsc}

A {\em string commitment\/} scheme is an extension of the well-studied
and powerful cryptographic primitive of {\em bit commitment\/}.  In
such schemes, one party, Alice, wishes to commit an entire string~$x
\in \{0,1\}^n$ to another party, Bob. The protocol is required to be
such that Bob should not be able to identify the string until it is
revealed by Alice.  In turn, Alice should not be able to renege on her
commitment at the time of revelation.  Formally, quantum string
commitment protocols are defined as
follows~\cite{BuhrmanCHLW06,Jain05}.
\begin{definition}[Quantum string commitment ($\QSC$)]
\label{def:QSC}
Let $P = \{p_x: x \in \{0,1\}^n \}$ be a probability distribution and
let $B$ be a measure of information contained in an ensemble of
quantum states. A $(n,a,b)$-$B$-$\QSC$ protocol for $P$ is a quantum
communication protocol between two parties, Alice and Bob.  Alice gets
an input $x \in \{0,1\}^n$ chosen according to the distribution
$P$. The starting joint state of the qubits of Alice and Bob is some
pure state independent of~$x$.  The protocol runs in two phases: the
commit phase, followed by the reveal phase.  There are no intermediate
measurements during the protocol. At the end of the reveal phase, Bob
measures his qubits according to a POVM~$\{M_y \;:\; y \in \{0,1\}^n\}
\cup \{I - \sum_y M_y\}$ to determine the value of the committed
string by Alice or to detect cheating.  The protocol satisfies the
following properties.
\begin{enumerate}

\item {\bf (Correctness)} Suppose Alice and Bob act honestly.  Let
$\rho_x$ be the state of Bob's qubits at the end of the reveal phase
of the protocol, when Alice gets input $x$. Then $(\forall x,y)\ \Tr\;
M_y \rho_x = 1$ iff $x=y$, and 0 otherwise.

\item {\bf (Concealing property)} Suppose Alice acts honestly, and Bob
possibly cheats, i.e., deviates from the protocol in his local
operations.  Let $\sigma_x$ be the state of Bob's qubits after the
commit phase when Alice gets input $x$. Then the $B$
information~$B(\E)$ of the ensemble $\E = \{p_x, \sigma_x \}$ is at
most $b$. In particular, this also holds when both Alice and Bob
follow the protocol honestly.

\item {\bf (Binding property)} Suppose Bob acts honestly , and Alice
possibly cheats. Let $c \in \{0,1\}^n$ be a string in a special
cheating register $C$ with Alice that she keeps independent of the
rest of the registers till the end of the commit phase. Let $\tau_c$
be the state of Bob's qubits at the end of the reveal phase when Alice
has $c$ in the cheating register. Let $q_c \eqdef \Tr\; M_c
\tau_c$. Then $$ \sum_{c \in \{0,1\}^n } p_c q_c \leq 2^{a - n}$$
\end{enumerate}
\end{definition}

The idea behind the above definition is as follows.  At the end of the
reveal phase of an honest run of the protocol Bob identifies $x$ from
$\rho_x$ by performing the POVM measurement $\{M_y\} \cup \{ I -
\sum_y M_y \}$. He accepts the committed string to be $x$ iff the
observed outcome~$y = x$; this happens with probability $\Tr\; M_x
\rho_x$. He declares that Alice is cheating if outcome $I - \sum_x
M_x$ is observed.  Thus, at the end of an honest run of the protocol,
with probability~$1$, Bob accepts the committed string as being
exactly Alice's input string.  The concealing property ensures that
the amount of $B$ information about~$x$ that a possibly cheating Bob
gets is bounded by $b$. In {\em bit\/}-commitment protocols, the
concealing property is quantified in terms of the probability with
which Bob can guess Alice's bit. Here we instead use different notions
of information contained in the corresponding ensemble.  The binding
property ensures that when a cheating Alice wishes to postpone
committing to a string string until after the commit phase, then she
succeeds in forcing an honest Bob to accept her choice with bounded
probability (in expectation).

{\em Strong\/} string commitment, in which both parameters~$a,b$ above
are required to be~$0$, is impossible for the same reason that of {\em
strong\/} bit-commitment protocols are
impossible~\cite{Mayers97,LoC97}. Weaker versions are nonetheless
possible, and exhibit a trade-off between the concealing and binding
properties.  The trade-off between the parameters~$a$ and~$b$ has been
studied by several researchers~\cite{Kent03,BuhrmanCHLW06,Jain05}.
Buhrman, Christandl, Hayden, Lo, and Wehner~\cite{BuhrmanCHLW06} study
this trade-off both in the scenario of a single execution of the
protocol and also in the asymptotic regime, with an unbounded number
of parallel executions of the protocol.  In the asymptotic scenario,
they show the following result in terms of Holevo information (which
is denoted by~$\chi$).

\begin{theorem}[\cite{BuhrmanCHLW06}]
\label{thm:harry}
Let $\Pi$ be an $(n, a_1, b)$-$\chi$-$\QSC$ scheme. Let $\Pi_m$ represent
$m$ parallel executions of $\Pi$ (so~$\Pi_1 = \Pi$).  Let $a_m$ represent the
binding parameter of $\Pi_m$ and let $a \eqdef \lim_{m \rightarrow
\infty} a_m/m$.  Then, $ a + b \geq  n $.
\end{theorem}

Jain~\cite{Jain05} shows a similar trade-off result regarding $\QSC$s,
in terms of the divergence information of an ensemble (denoted
by~$\rD$).
\begin{theorem}[\cite{Jain05}]
\label{thm:jain}
For single execution of the protocol of an $(n,a,b)$-$\rD$-$\QSC$ scheme, 
$$ a + b + 8 \sqrt{b + 1} + 16  \geq  n. $$
\end{theorem}

\suppress{ It was shown by Jain, Radhakrishnan and
Sen~\cite{jain:divrelsub, jain:substate} that for any two states
$\rho, \sigma$, $D(\rho \| \sigma) \leq S(\rho \| \sigma) + 1$, which
implies from Definitions~\ref{def:holevo} and~\ref{def:divinf} that }

As mentioned before, for any ensemble $\E$, divergence information is
bounded by the Holevo~$\chi$-information~$\rD(\E) \leq \chi(\E) + 1$.
This immediately implies:
\begin{theorem}[\cite{Jain05}]
\label{thm:jainchi}
For single execution of the protocol of a $(n,a,b)$-$\chi$-$\QSC$ scheme 
$$ a + b + 8 \sqrt{b + 2} +17  \geq  n. $$
\end{theorem}
As Jain shows, this implies the asymptotic result due to Buhrman {\it et
al.\/} (Theorem~\ref{thm:harry}). 

The separation that we demonstrate between divergence and Holevo
information (Theorem~\ref{thm-intro}) shows that for some ensembles
over~$n$ qubits, $\rD(\E)$ may be a~$\log n$ larger than~$\chi(\E)$.
For such ensembles the binding-concealing trade-off of
Theorem~\ref{thm:jain} is stronger than that of
Theorem~\ref{thm:harry}.

\suppress{
For example,
Theorem~\ref{thm-ensemble} shows that there exists an ensemble
where $\rD(\E) = \theta(n/\log n)$, and $\chi(\E) =
\Omega(n)$. For such an ensemble the lower bound on $a$ provided by
Theorem~\ref{thm:jain} is still quite significant whereas it could
possibly be very small as provided by Theorem~\ref{thm:harry}. 
}

\subsection{Privacy trade-off for two-party protocols for relations}

\newcommand{\setmem}{\mathsf{SetMemb}}

Let us consider two-party protocols between Alice and Bob for computing
a relation~$f \subseteq \cX \times \cY \times \cZ$. Jain,
Radhakrishnan, and Sen~\cite{JainRS02} studied to what extent the two
parties may solve~$f$ while keeping their respective inputs
hidden from the other party. They showed the following:
\begin{result}[\cite{JainRS05}, informal statement]
\label{res:div}
Let~$\mu$ be a product distribution on~$\cX \times \cY$. Let
$Q_{1/3}^{\mu, A \rightarrow B}(f)$ represent the one-way
distributional complexity of $f$ with a single communication from
Alice to Bob; and distributional error under $\mu$ at most $1/3$.
Let~$X$ and~$Y$ represent the random variables corresponding to Alice
and Bob's inputs respectively. If there is a quantum communication
protocol for $f$ where Bob {\em leaks} divergence
information at most~$b$ about his input $Y$, then Alice leaks divergence
information at least $\Omega(Q_{1/3}^{\mu, A \rightarrow
B}(f)/2^{O(b)})$ about her input $X$. Similar statement also holds
with the roles of Alice and Bob interchanged.
\end{result}
From the upper bound on the divergence information in terms of Holevo
information this immediately implies the following.
\begin{result}[\cite{JainRS05}, informal statement]
\label{res:holevo}
Let~$\mu$ be a product distribution on~$\cX \times \cY$. Let
$Q_{1/3}^{\mu, A \rightarrow B}(f)$ represent the one-way
distributional complexity of $f$ with a single communication from
Alice to Bob; and distributional error under $\mu$ at most $1/3$.
Let~$X$ and~$Y$ represent the random variables corresponding to Alice
and Bob's inputs respectively. If there is a quantum communication
protocol for $f$ where Bob {\em leaks} Holevo
information at most~$b$ about his input $Y$, then Alice leaks Holevo
information at least $\Omega(Q_{1/3}^{\mu, A \rightarrow
B}(f)/2^{O(b)})$ about her input $X$. Similar statement also holds
with the roles of Alice and Bob interchanged.
\end{result}
It follows from Theorem~\ref{thm-intro} that Result~\ref{res:div} is
much stronger than the second, Result~\ref{res:holevo} in case the
ensembles arising in the protocol between Alice and Bob has
divergence information much smaller than its Holevo information.

\end{document}